\documentclass[10pt, conference]{IEEEtran}
\IEEEoverridecommandlockouts
\usepackage[noadjust]{cite}
\usepackage{amsmath, amssymb, amsfonts, url}
\usepackage{algorithmic}
\usepackage{graphicx, tabularx, booktabs, subcaption}
\usepackage{textcomp}
\usepackage{xcolor}
\usepackage[acronym]{glossaries}

\def\BibTeX{{\rm B\kern-.05em{\sc i\kern-.025em b}\kern-.08em
    T\kern-.1667em\lower.7ex\hbox{E}\kern-.125emX}}

\begin{document}

\title{Orbit-Aware Split Learning: Optimizing LEO Satellite Networks for Distributed Online Learning
\thanks{This work is part of the project SOFIA PID2023-147305OB-C32 funded by MICIU/AEI/10.13039/501100011033 and FEDER/UE.}
}

\author{\IEEEauthorblockN{Marc Martinez-Gost\IEEEauthorrefmark{1}\IEEEauthorrefmark{2}, Ana Pérez-Neira\IEEEauthorrefmark{1}\IEEEauthorrefmark{2}\IEEEauthorrefmark{3}}
\IEEEauthorblockA{
\IEEEauthorrefmark{1}Centre Tecnològic de Telecomunicacions de Catalunya, Spain\\
\IEEEauthorrefmark{2}Dept. of Signal Theory and Communications, Universitat Politècnica de Catalunya, Spain\\
\IEEEauthorrefmark{3}ICREA Acadèmia, Spain\\
\{mmartinez, aperez\}@cttc.es
}}

\newacronym{AI}{AI}{Artificial Intelligence}
\newacronym{FL}{FL}{federated learning}
\newacronym{FLOPS}{FLOPS}{floating point operations per second}
\newacronym{FSPL}{FSPL}{free space path loss}
\newacronym{GS}{GS}{ground station}
\newacronym{IoT}{IoT}{Internet of Things}
\newacronym{ISL}{ISL}{inter-satellite link}
\newacronym{LEO}{LEO}{low Earth orbit}
\newacronym{Sat-IoT}{Sat-IoT}{satellite Internet of Things}
\newacronym{SL}{SL}{split learning}
\newacronym{SMEC}{SMEC}{satellite mobile edge computing}

\maketitle
\begin{abstract}
This paper proposes a split learning (SL) framework tailored for Low Earth Orbit (LEO) satellite constellations, leveraging their cyclical movement to improve energy efficiency. Although existing research focuses on offloading tasks to the non-terrestrial network (NTN) infrastructure, these approaches overlook the dynamic movement patterns of LEO satellites that can be used to efficiently distribute the learning task. In this work, we analyze how LEO satellites, from the perspective of ground terminals, can participate in a time-window-based model training. By splitting the model between a LEO and a ground terminal, the computational burden on the satellite segment is reduced, while each LEO satellite offloads the partially trained model to the next satellite in the constellation. This cyclical training process allows larger and more energy-intensive models to be deployed and trained across multiple LEO satellites, despite their limited energy resources. We formulate an optimization problem that manages radio and processing resources, ensuring the entire data is processed during each satellite pass while minimizing the energy consumption. Results demonstrate that the proposed architecture optimizes both communication and processing resources, achieving up to 97\% energy savings compared to direct raw data transmission. This approach offers a more scalable and energy-efficient way to train complex models, enhancing the capabilities of LEO satellite constellations in Artificial Intelligence-driven applications.
\end{abstract}

\section{Introduction}
The last two decades have seen an unprecedented growth towards the deployment of \gls{SMEC} over \gls{LEO} satellite constellations \cite{smec}. \gls{SMEC} extends traditional MEC concepts to satellite networks and comprises exploiting the computational resources (e.g., servers, GPUs, or specialized processing units) on satellites to perform tasks like data processing and caching.
This trend can be attributed to several key technological and economic advancements.
Specifically, modern satellites are now equipped with advanced capabilities, integrating cutting-edge communication, computing, and sensing technologies. These enhancements allow \gls{LEO} satellites not only to provide traditional services, such as global communication and Earth observation, but also enable them to support more sophisticated applications, including distributed computing and \gls{AI} model inference closer to the data source or end users.

Despite the significant potential of \gls{SMEC} in enabling innovative satellite-assisted services, current research has predominantly focused on offloading tasks from ground terminals to \gls{LEO} satellites. In contrast, the reverse direction—offloading tasks from satellites to ground stations—has received far less attention \cite{Denby2020, marc2022, israel2024}. This paradigm, termed Native \gls{SMEC} or \gls{Sat-IoT}  \cite{satiot}, arises because the data originates at the satellite premises due to their advanced sensing capabilities, generating vast amounts of data that must be processed in a timely manner. However, this is challenging due to the limited computational and energy resources onboard satellites. At the same time, transmitting all raw data to the ground is unfeasible given the constrained bandwidth available in satellite communication links. These challenges underscore the need for efficient mechanisms to balance on-board processing and ground-based offloading in Native \gls{SMEC}.

The Native \gls{SMEC} environment inherently calls for distributed solutions and online learning approaches, as data is generated and distributed across the entire satellite constellation while conditions and data evolve dynamically over time.
In this work, we propose a \gls{SL} architecture that divides the \gls{AI} model processing between ground terminals and \gls{LEO} satellites. The architecture leverages the orbital dynamics of \gls{LEO} satellites, enabling tasks to be processed sequentially as satellites pass over the ground station. This approach ensures that the data captured by all satellites in the orbital ring contributes to the model training, facilitating a cyclical, distributed training process. By transferring intermediate representations instead of raw data, the proposed \gls{SL} framework reduces communication overhead and supports energy-efficient operations, while exploiting the natural movement of \gls{LEO} satellites to harness the rich, distributed data collected across the constellation. This enables the deployment of large-scale \gls{AI} models and ensures scalable, real-time adaptation to the unique challenges of Native \gls{SMEC}.

The proposed architecture offers additional benefits beyond enabling distributed online learning. Specifically, we formulate an energy minimization problem to efficiently distribute the workload between the ground terminal and the \gls{LEO} satellite. This involves selecting the optimal communication and computational resources, all while adhering to the timing constraints imposed by the orbital dynamics of the constellation. Unlike classical \gls{SMEC} approaches, which often involve centralized and computationally intensive solutions, the proposed architecture results in a distributed optimization problem that is much easier to solve. This simplification arises from the decoupling of optimization variables, which ensures scalability and feasibility in real-time applications while maintaining energy efficiency and meeting system constraints.

The remaining part of the paper proceeds as follows:
Section II discusses the deployment of edge learning over \gls{LEO} satellite networks; Section III presents the system model and Section IV introduces the resource optimization problem.
Section V analyses the performance of \gls{SL} over \gls{LEO} constellations in different learning tasks and section VI concludes the paper.

\section{Edge Learning over \gls{LEO} networks}

The integration of \gls{AI} services within \gls{SMEC}, and particularly in Native \gls{SMEC}, is essential for addressing the unique needs and specifications of satellite-driven applications.
\begin{itemize}
    \item \textbf{Efficient data processing}: The vast amounts of data generated by satellites, such as high-resolution imagery, video, and sensor measurements, demand advanced processing techniques capable of extracting actionable insights efficiently and in a timely manner. \gls{AI} methods, such as deep learning and computer vision, are well-suited for tasks like object detection, anomaly recognition, and predictive analytics, making them indispensable for transforming raw data into useful information.

    \item \textbf{Autonomous and adaptive intelligence}: The deployment of \gls{AI} services supports the development of next-generation satellite capabilities, including autonomous operation and on-orbit intelligence. This allows satellites to not only sense their environment but also make context-aware decisions, reducing dependence on constant communication with ground stations. Online learning  \cite{hoi2021online}, for instance, is a key technique that enables continuous adaptation to changing conditions, ensuring real-time, intelligent decision-making onboard satellites.
\end{itemize}

By meeting the growing demand for intelligent processing in space, \gls{AI} services are poised to become a cornerstone of \gls{SMEC} applications. Nonetheless, the Native \gls{SMEC} paradigm imposes a unique set of constraints on the design of \gls{AI} models, tailored specifically to the satellite environment. Unlike terrestrial systems, where centralized data processing or ample computational resources are often available, the satellite context requires \gls{AI} models to operate under stringent limitations while addressing the distinct characteristics of distributed satellite systems.  
\begin{itemize}
    \item \textbf{Distributed learning:} \gls{AI} models in Native \gls{SMEC} must inherently be distributed to align with the natural distribution of data among satellites. Each satellite captures localized data, which are geographically and temporally segmented. Consequently, \gls{AI} algorithms must be designed to process and aggregate insights collaboratively across multiple satellites without relying on centralized datasets. This requires the use of distributed learning techniques to ensure effective model training and inference across the constellation while minimizing inter-satellite communication overhead.

    \item \textbf{Energy-efficient \gls{AI} design}: The energy consumption of \gls{AI} models is a critical design factor. Satellites are powered by solar energy and operate under tight energy budgets, making it essential to develop lightweight algorithms that can perform complex tasks without exhausting power resources. Techniques like edge \gls{AI} inference and real-time optimizations become critical in this context.
\end{itemize}

\Gls{FL} \cite{FL1, FL2, FL3} has emerged as a promising technique for distributed \gls{AI} model training in Native \gls{SMEC}, particularly due to its ability to handle the decentralized nature of satellite data. In \gls{FL}, instead of sending raw data to a central server, satellites train models locally and share only updates, preserving data privacy and reducing the need for large-scale data transmissions. However, \gls{FL} presents several challenges in the satellite context. First, communication overhead for aggregating updates can consume significant bandwidth and energy. As satellite constellations grow, the communication burden increases, potentially leading to inefficiencies. Second, while \gls{FL} distributes training across multiple entities, it may still require large models to achieve high performance for complex tasks. Such large models are not energy-preserving, as training and updating them on satellites with constrained power resources can be prohibitively expensive. Thus, while \gls{FL} is well-suited for distributed \gls{AI} in satellite networks, addressing communication, computational, and energy constraints is critical to its success in this context.

\Gls{SL} \cite{SL} is another paradigm for distributed learning that addresses some of the challenges of \gls{FL}. In \gls{SL}, the training process is divided into two parts: one part is computed locally on the edge device (e.g., satellite), while the other part is computed on a central server (e.g., ground station). Instead of sharing the entire model or raw data, \gls{SL} only transmits intermediate outputs (activations) between the split parts, significantly reducing communication overhead. This approach allows for the deployment of larger, more computationally demanding models while mitigating the communication and energy constraints inherent in satellite networks.
Nonetheless, the literature on \gls{SL} over satellite networks is very limited \cite{SL2, SL3, SL4}.

The novel idea in this work is that the dynamics of the satellite constellation offer a unique opportunity to implement \gls{SL} in a way that aligns naturally with the orbital movement of the satellites. 
The algorithm is initially split between the satellite and the ground terminal. After the satellite pass, the next incoming satellite can continue the training by processing the data, taking advantage of the satellite constellation's cyclical movement.
This architecture offers several key benefits, which are:
\begin{itemize}
    \item \textbf{Support for heterogeneous devices}: energy-constrained satellites may skip the learning, bypassing their limitations and ensuring efficient use of energy resources.

    \item \textbf{Enabling large \gls{AI} models}: \gls{SL} facilitates the deployment of large AI models that would otherwise exceed the satellite's computational capacity, enabling advanced applications.
    
    \item \textbf{Reduced ground infrastructure requirements}: Reducing the processing burden is particularly valuable for enabling \gls{AI} in remote and underserved regions. 

    \item \textbf{Scalable and efficient solutions}: the ability of \gls{SL} to operate with minimal reliance on centralized infrastructure aligns well with the distributed nature of Native \gls{SMEC}, offering a scalable, efficient solution for integrating \gls{AI} into satellite systems.
\end{itemize}

\section{System Model}
\begin{figure}[t]
\centering
\includegraphics[width=\columnwidth]{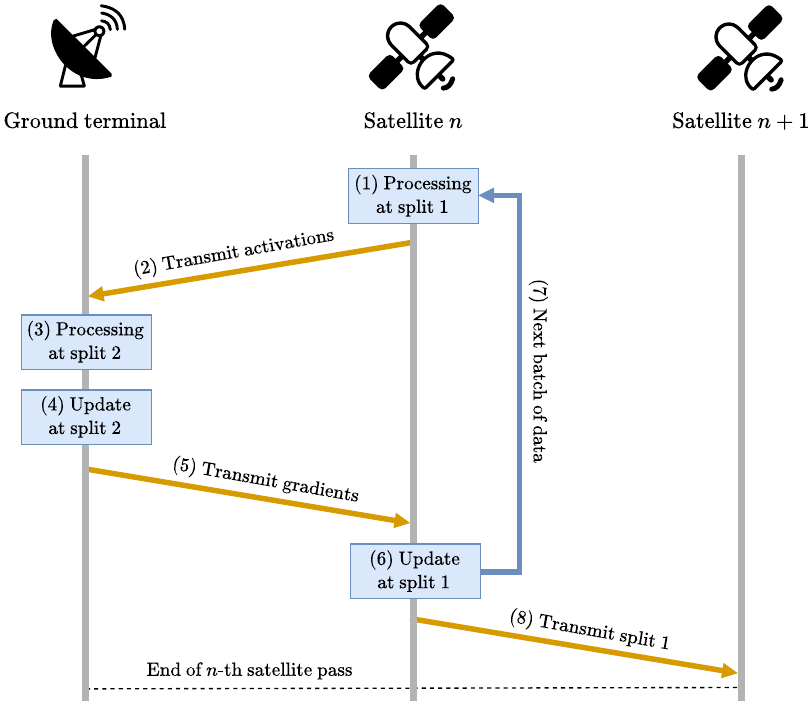}
\caption{Procedure of the proposed \gls{SL} architecture. After step (8), the whole process is repeated between $(n+1)$-th satellite and the ground terminal.}
\label{fig:SL_procedure}
\vspace{0 pt}
\end{figure}

We consider a set of \gls{LEO} satellites capturing data with its sensors (e.g., optical images, hyperspectral data, SAR, etc.) with the goal of training an algorithm (e.g, a neural network).
Due to the overall amounts of data and the limited communication bandwidth, sending the raw data towards the ground for processing is not feasible.
On the other hand, running the algorithm at the satellite premises is complex because it is energy-intensive and the data is distributed across the \gls{LEO} constellation.
Following the \gls{SL} paradigm, the algorithm is split into two parts: the first piece is deployed at the \gls{LEO} and the latter at the ground terminal.
During the satellite pass of a specific \gls{LEO}, the satellite and the ground terminal collaboratively execute the algorithm, leveraging their communication and computational capabilities. Once the satellite pass concludes, the satellite hands off the split to the following satellite in the orbital ring. This succeeding satellite will continue the training process using its local data.
Figure \ref{fig:SL_procedure} illustrates this joint communication and processing procedure.
The proposed scenario can be formulated in the opposite direction, this is, a ground terminal with stringent energy constraints exploiting the communication capabilities with a set of \gls{LEO} satellites. 

\subsection{Network architecture}
For the sake of simplicity we assume the ground terminal communicates with a single \gls{LEO} satellite during the satellite pass. All satellites belong to the same orbital ring, which is comprised of $N$ evenly distributed satellites deployed at altitude $h$. We will determine the duration of each satellite pass over the ground terminal, as well as the distances for the ground-to-satellite link (used for \gls{SL} communication) and the inter-satellite links (used to exchange the satellite's portion of the model). These metrics are essential for understanding the communication and timing constraints of the proposed system.

The period of the \gls{LEO} satellite is
\begin{equation}
    T_o = \sqrt{\frac
    {4\pi^2~(R_E+h)^3}
    {GM}},
    \label{eq:period}
\end{equation}
where $R_E$ and $M$ are the radius and the mass of the Earth, respectively; $G$ is the universal gravitational constant \cite{leyva2022ngso}. The distance of the ground to satellite link is
\begin{equation}
    d(\varepsilon) = \sqrt{R_E^2\sin^2(\varepsilon)+
    2R_Eh+h^2} - 
    R_E\sin(\varepsilon),
    \label{eq:distance}
\end{equation}
where $\varepsilon$ is the elevation angle. The maximum distance is achieved with the minimum elevation angle, $\varepsilon_{min}$, which determines the start of the satellite pass.

The Earth central angle determines the shift in the position of the device with respect to the satellite’s nadir point. With a minimum elevation angle, this determines the angular distance of the satellite pass:
\begin{align}
    \alpha_{\text{pass}} = 2\arccos\left(
    \frac
    {(R_E+h)^2 + R_E^2-d(\varepsilon_{min})^2}
    {2\left(R_E^2+R_Eh
    \right)}
    \right)
    \label{eq:angle_pass}
\end{align}

Finally, the duration of the satellite pass can be computed as
\begin{equation}
    T_{\text{pass}} =
    \frac{T_o~\alpha_{\text{pass}}}{\pi}
    \label{eq:T_pass}
\end{equation}
The distance between consecutive satellites in the orbital plane can be considered a constant:
\begin{equation}
    d_{\text{ISL}} =
    2(R_E+h)\sin(\pi/N)
    \label{eq:d_isl}
\end{equation}

\begin{figure}[t]
\centering
\includegraphics[width=\columnwidth]{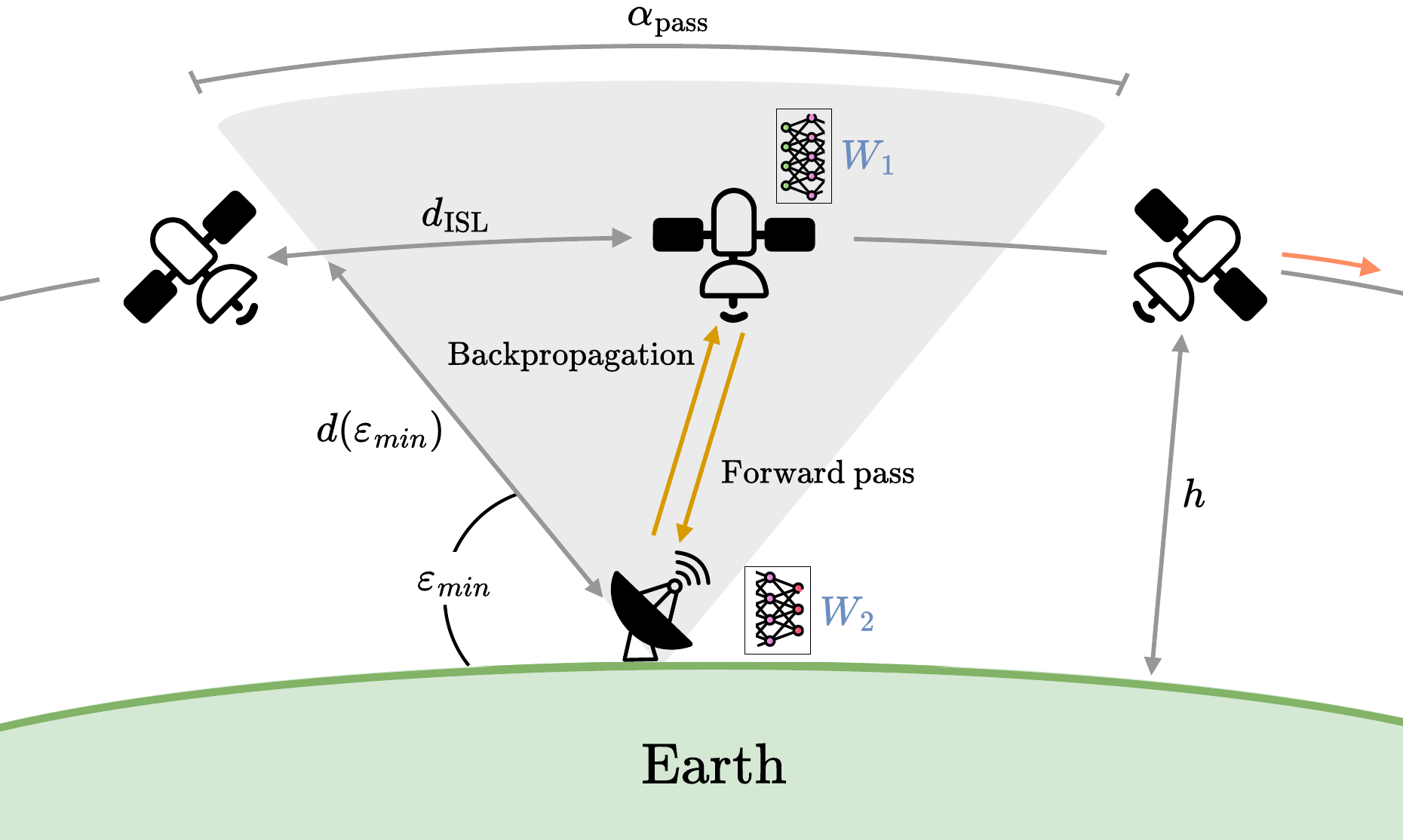}
\caption{\gls{SL} architecture deployed over a ring of \gls{LEO} satellites and a ground terminal. The algorithm is exemplified with a neural network, including the forward and backward passes.}
\label{fig:system_model}
\vspace{0 pt}
\end{figure}

In the end, the ground terminal has visibility of each \gls{LEO} for $T_{\text{pass}}$ seconds. After that slot, the succeeding satellite takes the task. Figure \ref{fig:system_model} illustrates the network architecture.

\subsection{Computation}
Let $W$ represent the computational work required by the algorithm of interest (e.g., training a neural network), expressed as the number of \gls{FLOPS} needed to execute the algorithm on a single unit of data (e.g., an image).
The processing time is a random variable influenced by task allocation inefficiencies and memory or data management overhead \cite{gpu}. We assume a perfect deterministic model, which accounts for most of the processing time. We define the processing time as
\begin{equation}
    T_{\text{proc}}(W,f_{p}) = \frac{D~W}{N_c~N_{\text{FLOPS}}~f_{p}},
    \label{eq:t_proc}
\end{equation}
where $D$ is the input size (e.g., pixels), $N_c$ is the number of cores of the processor, $N_{\text{FLOPS}}$ is the number of \gls{FLOPS} executed per CPU cycle and $f_{p}$ is the frequency of the processor.

The corresponding energy consumption can be then computed as
\begin{equation}
    E_{\text{proc}}(W,f_{p}) = 
    P(f_{p})T_{\text{proc}}(W,f_{p})=
    \frac{D~W~P_{p}~f_{p}^2}{N_c~N_{\text{FLOPS}}~f_{p, \text{max}}^3},
    \label{eq:E_proc}
\end{equation}
where $f_{p, \text{max}}$ is the maximum frequency of operation and $P_p$ is the power at maximum frequency.

According to Figure \ref{fig:system_model}, the task is split between the ground terminal and a single \gls{LEO}. We assume a sequential model (e.g., a neural network), which can be easily split in two parts. Each part carries a work of $W_i(\ell)$, for $i=\{1,2\}$, where $\ell$ corresponds to the splitting point of the model (e.g., layer in the neural architecture). Thus, we have $W=W_1(\ell)+W_2(\ell)$. Without loss of generality, we assume that the first split is held at the satellite, while the second one at the ground terminal. Thus, the corresponding processing time and energy consumption for each split are computed by substituting $W_i$ in \eqref{eq:t_proc} and \eqref{eq:E_proc}, respectively.

\subsection{Communication}
The output of the first split needs to be conveyed to the second split to complete the processing. We define $D_{\text{tx}}$ as the amount of data to be transmitted and define the associated communication time as
\begin{equation}
    T_{\text{comm}}(p_{\text{tx}}) = 
    \frac{D_{\text{tx}}(\ell)}{R(p_\text{tx})} =
    \frac{D_{\text{tx}}(\ell)}
    {B\log_2\left(
    1+\frac{p_{\text{tx}}~G}{\text{FSPL}~\sigma^2}
    \right)}
    \label{eq:t_comm},
\end{equation}
where $R$ is the communication rate and depends on the transmission power $p_{\text{tx}}$, allocated bandwidth $B$, overall antenna gain $G$, \gls{FSPL} and channel noise $\sigma^2$. Notice that the amount of transmitted data depends on the split point. 

The corresponding energy devoted to communication can be computed as
\begin{equation}
    E_{\text{comm}}(p_{\text{tx}}) = 
    p_{\text{tx}}~T_{\text{comm}}(p_{\text{tx}})=
    \frac{p_{\text{tx}}D_{\text{tx}}(\ell)}
    {B\log_2\left(
    1+\frac{p_{\text{tx}}~G}{\text{FSPL}~\sigma^2}
    \right)}
    \label{eq:e_comm}
\end{equation}
and the average propagation time corresponds to $T_{\text{prop}}=\bar{d}/c$, where $\bar{d}$ is the average distance during satellite pass and $c$ is the speed of light.

Finally, to continue the learning, it is necessary that the current \gls{LEO} sends the corresponding split to the succeeding satellite before the satellite pass is over. The communication is established with an \gls{ISL} and the devoted time is 
\begin{equation}
    T_{\text{ISL}} = 
    \frac{D_{\text{ISL}}(\ell)}
    {R_{\text{ISL}}},
\end{equation}
where $D_{\text{ISL}}$ corresponds to the size of the split in bits and $R_{\text{ISL}}$ is the \gls{ISL} data rate. The associated propagation time is $d_{\text{ISL}}/{c}$ and the spent energy is
$E_{\text{ISL}}=p_{\text{tx}}^{\text{ISL}}~T_{\text{ISL}}$, where $d_{\text{ISL}}$ is the \gls{ISL} distance and $p_{\text{tx}}^{\text{ISL}}$ is the power used in to transmit in the \gls{ISL}.

\section{Energy Minimization in \gls{SL}}

We assume that a batch of data $D$ is to be processed by every LEO. 
We assume that the first split is deployed at the satellite, but the problem is mathematically equivalent if formatted in the opposite direction.

The total energy spent by the system during a given satellite pass is
\begin{align}
    E_{\text{total}} =& 
    ~E_{\text{proc}}(W_1,f_{p}^{\text{LEO}}) +
    E_{\text{comm}}(p_{\text{tx}}^{\text{LEO}}) +
    E_{\text{proc}}(W_2,f_{p}^{\text{GS}})\nonumber\\
    &+
    E_{\text{comm}}(p_{\text{tx}}^{\text{GS}}) +
    E_{\text{ISL}}(p_{\text{tx}}^{\text{ISL}}),
    \label{eq:e_total}
\end{align}
where $f_{p}^{\text{LEO}}$ and $f_{p}^{\text{GS}}$ are the frequency of the processor at the satellite and ground terminal, respectively; $p_{\text{tx}}^{\text{LEO}}$ and $p_{\text{tx}}^{\text{GS}}$ are the corresponding transmission powers. 

The overall latency per satellite pass is
\begin{align}
    T_{\text{total}} =& 
    ~T_{\text{proc}}(W_1,f_{p}^{\text{LEO}}) +
    T_{\text{comm}}(p_{\text{tx}}^{\text{LEO}}) + 
    2~T_{\text{prop}} \nonumber\\
    &+
    T_{\text{proc}}(W_2,f_{p}^{\text{GS}}) +
    T_{\text{comm}}(p_{\text{tx}}^{\text{GS}}) +
    T_{\text{ISL}},
    \label{eq:t_total}
\end{align}
where $T_{\text{prop}}$ occurs twice because of the forward propagation of the data and the backward propagation of the gradient updates.

We formulate an optimization problem to minimize the overall energy per satellite pass by choosing the frequency of the processor and the transmission power at both devices:
\begin{subequations}
\begin{align}
\nonumber
& \underset{f_{p}^{\text{GS}}, f_{p}^{\text{LEO}}, p_{\text{tx}}^{\text{GS}}, p_{\text{tx}}^{\text{LEO}}}{\text{minimize}} ~~~~
E_{\text{total}} \\
& \hspace{-10pt}\qquad\text{subject to}~~~~~~~
T_{\text{total}} \leq T_{\text{pass}}\label{const:latency}\\
& ~~~~~~~~~~~~~~~~~~~~~
f_{p}^{\text{m}}\leq f_{p, \text{max}}^{\text{m}},~~~~\text{m}=\{\text{GS, LEO}\}\label{const:frequency}\\
& ~~~~~~~~~~~~~~~~~~~~~
p_{\text{tx}}^{\text{m}}\leq P_{\text{tx}, \text{max}}^{\text{m}},~~~~\text{m}=\{\text{GS, LEO}\}\label{const:power}
\end{align}
\label{eq:min_energy}
\end{subequations}

Constraint \eqref{const:latency} ensures that the data batch is processed before the satellite pass is finished and \eqref{const:frequency} prevents the processor from operating beyond the maximum clock frequency in both splits. Constraint \eqref{const:power} ensures that the transmission power does not exceed the power budget in both uplink ($\text{m}=\text{GS}$) and downlink ($\text{m}=\text{LEO}$). Likewise, this constraint ensures that the maximum power does not exceed the maximum achievable rate.

Problem \eqref{eq:min_energy} is quasiconvex, because the cost is a quasiconvex function and the constraints are convex. This problem can be easily solved with the bisection method and ensure convergence to an optimal minimum \cite{Boyd}. By limiting the optimization to the interaction between a single \gls{LEO} and the ground terminal, the system avoids the complexities of global optimization across the entire constellation. This makes the system inherently scalable to larger satellite constellations without a significant increase in computational with no added computational burden for coordination.
Furthermore, the time-window-based offloading design reduces the complexity of synchronization across the system.


\section{Results}
\label{sec:results}
In this section, we evaluate the proposed architecture using two distinct image processing tasks: image compression and image classification. The first task demonstrates the advantages of splitting the algorithm, while the second highlights the impact of the split point within the algorithm.
Table \ref{tab:parameters} provides the details of the simulation setup, including the constellation design, communication and computing parameters, as well as the dataset captured at the satellite premises. With this configuration, $T_\text{pass}\approx3.8$ minutes.

\begin{table}[t]
\renewcommand{\arraystretch}{1.4}
\centering
\caption{Parameter settings for performance evaluation.}
\begin{tabularx}{\columnwidth}{@{}lXll@{}}
\toprule
\multicolumn{2}{@{}l}{\textbf{Parameter}} & \textbf{Symbol} & \textbf{Settings}\\ 
\midrule

\multicolumn{2}{@{}l}{\textbf{Constellation}}\\
& Number of satellites per orbital plane & $N$ & $25$\\
& Altitude of deployment & $h$ & $550$~km\\
& Minimum elevation angle & $\varepsilon_{min}$ & $30^\circ$\\

\multicolumn{2}{@{}l}{\textbf{Communication}}\\
 & Maximum transmission power for the satellite-to-ground links & $P_{\text{tx, max}}$ & $10$~W\\
 & Bandwidth for satellite-to-ground links & $B$ & $500$~MHz\\
 & Transmission frequency for satellite-to-ground links & $f_{\text{tx}}$ & $20$~GHz\\
 & Transmission power in the intra-plane \gls{ISL} & $p_\text{tx}^\text{ISL}$ & $0.5$~W\\
 & Intra-plane \gls{ISL} data rate & $R_\text{ISL}$ & $5$ Gbps\\
 & Noise power & $\sigma^2_\text{dB}$& $-119$~dBW\\
 & Total antenna gain for the satellite-to-ground links & $G$ & $66.33$~dBi\\

 \multicolumn{2}{@{}l}{\textbf{Computing}}\\
 & Power consumption & $P_p$ & $15$~W\\
 & Maximum frequency clock & $f_{p,\text{max}}$ & $625$~MHz\\
 & Number of cores & $N_{\text{c}}$ & $1024$\\
 & Number of FLOPS & $N_{\text{FLOPS}}$ & $2$\\

 \multicolumn{2}{@{}l}{\textbf{Dataset}}\\
& Number of images in the training dataset per satellite &$N_\text{img}^\text{train}$ & $400$\\
& Average image size & $D$ & $1.605$ Mbits\\

 \bottomrule
\end{tabularx}
\label{tab:parameters}
\end{table}

\subsection{Image compression}
Image compression of satellite imagery is essential due to the limited resources available in satellite communication systems. As direct downloading of high-resolution images to the ground is often infeasible, compression reduces the data size, making it more manageable for transmission while minimizing energy consumption \cite{israel2024}.
To address the image compression task, we will implement an autoencoder neural network. An autoencoder consists of two main components: the encoder and the decoder. The encoder maps the input image into a lower-dimensional latent space, effectively compressing the data, while the decoder reconstructs the image from this compressed representation. This architecture enables the model to learn an efficient, task-specific compression scheme, potentially achieving compression ratios that exceed those of traditional algorithms like JPEG.

For this implementation, we use an autoencoder to compress color images from $224\times224$ pixels down to a $7\times7$ latent space. The encoder is allocated at the satellite premises, while the decoder is located at the ground terminal. Assuming 32 bits per datum, the LEO satellite transmits $D_\text{tx}=4.7$~kbits per image (with the same size assumed for the gradients in the uplink). Additionally, the encoder weighs $D_\text{ISL}=168.8$~kbits. We use the fvcore library \cite{fvcore} to estimate the load of the algorithm. The encoder and decoder require $W_1\approx302$ GFLOPS and $W_2\approx39$ MFLOPS, respectively.

\begin{figure}[t]
    \centering
    
    \begin{subfigure}[b]{\columnwidth}
        \centering
        \includegraphics[width=\columnwidth]{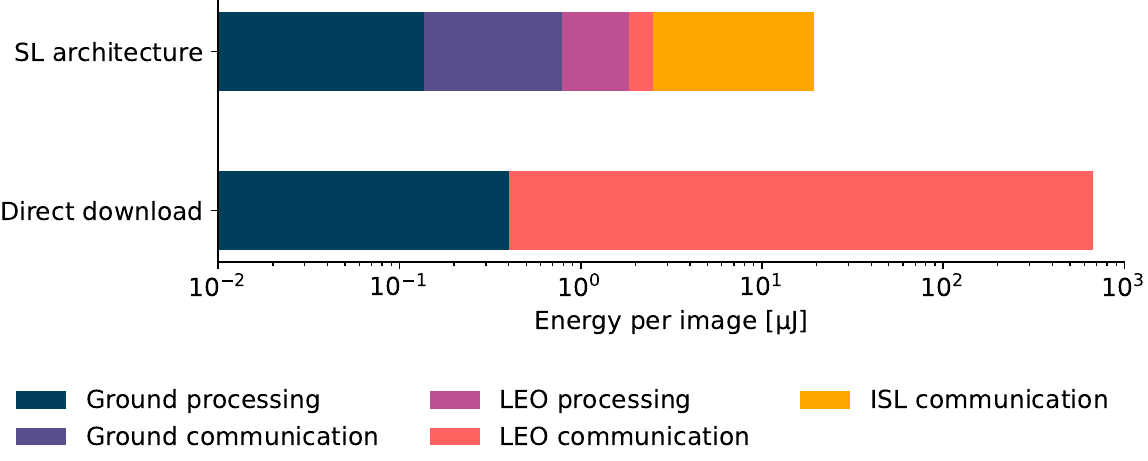}
        \label{fig:subfig1}
    \end{subfigure}
    
    \vspace{-0.2cm}

    \begin{subfigure}[b]{\columnwidth}
        \centering
        \includegraphics[width=\columnwidth]{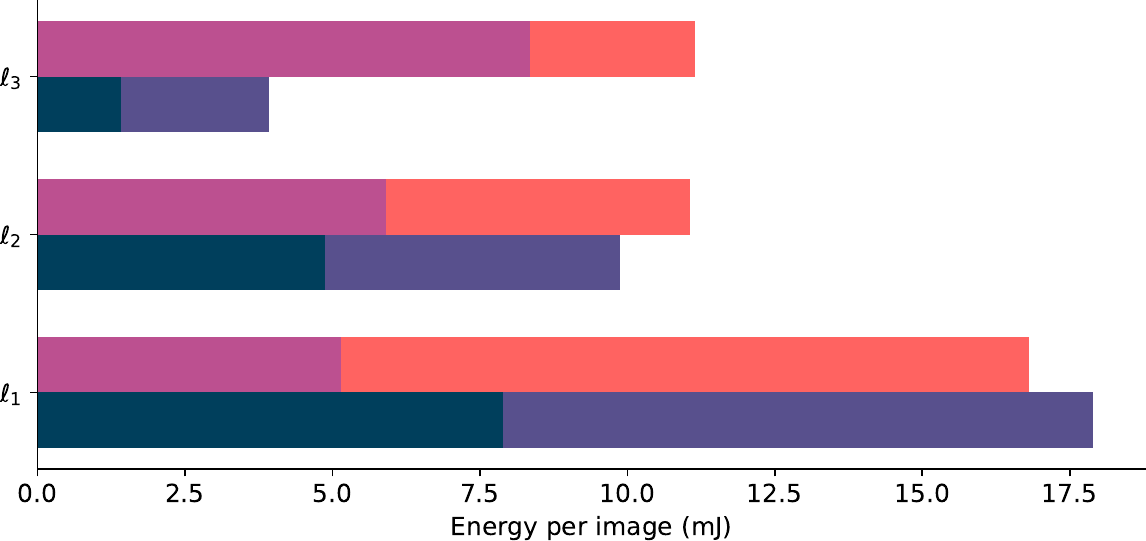}
        \label{fig:subfig2}
        \vspace{-0.4cm}
    \end{subfigure}
    \caption{Energy consumption: (Top) Autoencoder with \gls{SL} and direct download; (Bottom) ResNet-18 at three different splitting points $\ell$.}
    \label{fig:results}
\end{figure}

Figure \ref{fig:results} (Top) illustrates the energy spent during the online learning process of the autoencoder, decoupled into communication and processing. We compare the proposed \gls{SL} architecture with a direct download of raw data approach, where images are transmitted to the ground terminal, and the entire algorithm is executed on the ground device. In both scenarios, the optimization problem \eqref{eq:min_energy} is solved to minimize the overall energy expenditure of the system.
In the direct download approach, the energy consumption is high due to the communication overhead required to transmit high-resolution images. This approach does not face significant energy challenges in terms of maintaining latency constraints. 
On the other hand, the \gls{SL} approach significantly reduces energy consumption. By transmitting only the compressed latent representation rather than the raw image, the communication burden is reduced. While some energy is still needed for transmitting gradients and moving the encoder to the next satellite, the overall energy savings add up to 97$\%$ compared to the direct download method.

\subsection{Image classification}
Image classification of satellite imagery is crucial for applications where quick and accurate analysis of large-scale images is essential for monitoring vast areas. In such scenarios, \gls{SL} provides a key advantage by preserving the privacy of the data, as raw images are never transmitted. 
We use a ResNet-18 architecture, which is a lightweight and widely-used neural network for classification tasks  \cite{resnet}. Despite being efficient, the overall computational cost of the ResNet model is higher compared to the autoencoder used in the previous task, due to the complexity of the classification task and the deeper layers involved. The work of the algorithm and data sizes for three different splitting points $\ell$ are listed in Table \ref{tab:resnet}.

Figure \ref{fig:results} (Bottom) illustrates how varying the splitting point within the ResNet-18 architecture affects the overall energy consumption of the system. 
For this task, splitting the model to process more at the satellite premises (i.e., $\ell_3$) proves to be advantageous, as the satellite's output is smaller, reducing the communication cost and, thus, minimizing the energy spent on communication and ground processing.
These results show that the optimal splitting point depends on the learning model architecture, demanding tailored solutions for each use case.

\begin{table}[t]
\renewcommand{\arraystretch}{1.4}
\centering
\caption{Resource requirements for different splitting points in ResNet-18 architecture.}
\begin{tabularx}{\columnwidth}{@{}c c c c c@{}}
\toprule
 & $W_1$ (GFLOPS) & $W_2$ (GFLOPS) & $D_\text{tx}$ (Mbits) & $D_\text{ISL}$ (Mbits) \\ 
\midrule
$\ell_1$ & 1.765 & 3.714 & 6.423 & 369.056 \\
$\ell_2$ & 3.006 & 2.474 & 3.211 & 352.224 \\
$\ell_3$ & 4.243 & 1.237 & 1.605 & 285.024 \\
\bottomrule
\end{tabularx}
\label{tab:resnet}
\end{table}

\section{Conclusion}
This paper proposed a \gls{SL} architecture tailored for \gls{LEO} satellite constellations, leveraging their cyclical movement to efficiently distribute model training between satellites and ground terminals. By splitting models, the approach reduces computational burden on energy-constrained satellites while enabling the deployment of larger, energy-intensive models. Results demonstrated that the architecture minimizes overall energy consumption by optimizing communication and processing resources, achieving up to 97$\%$ energy savings compared to direct raw data transmission.
Future work will focus on investigating the optimal selection of key parameters for different satellite formations, considering their specific resource constraints.


\bibliographystyle{IEEEbib}
\bibliography{refs}

\end{document}